\documentclass[10pt]{iopart}
\usepackage{iopams,natbib,graphicx}

\newcommand{\beq}{\begin{equation}}
\newcommand{\eeq}{\end{equation}}
\newcommand{\bea}{\begin{eqnarray}}
\newcommand{\eea}{\end{eqnarray}}
\renewcommand{\cite}[1]{\citep{#1}}
\newcommand{\la}{\lesssim}
\newcommand{\ga}{\gtrsim}
\newcommand{\apj}{{\it Astrophys. J.}}
\newcommand{\apjs}{{\it Astrophys. J. Suppl.}}
\newcommand{\apjl}{{\it Astrophys. J. Lett.}}
\newcommand{\mnras}{{\it Mon. Not. R. Astron. Soc.}}
\newcommand{\prl}{{\it Phys. Rev. Lett.}}
\newcommand{\sun}{\odot}
\newcommand{\rem}[1]{ }

\bibpunct[,]{(}{)}{;}{a}{}{,}

\begin{document}

\title[Radiative cooling in relativistic shocks]{Radiative cooling in relativistic collisionless shocks. 
Can simulations and experiments probe relevant GRB physics?}

\author{Mikhail V Medvedev$^1$\footnote{Also: Institute for Nuclear Fusion, RRC ``Kurchatov
Institute", Moscow 123182, Russia}
and Anatoly Spitkovsky$^2$}

\address{$^1$ Department of Physics and Astronomy, University of Kansas, KS 66045}
\address{$^2$ Department of Astrophysical Sciences, Peyton Hall, Princeton
University, Princeton, NJ 08544}
\ead{medvedev@ku.edu}

\begin{abstract}
We address the question of whether numerical particle-in-cell (PIC) 
simulations and laboratory laser-plasma experiments can
(or will be able to, in the near future) model realistic 
gamma-ray burst (GRB) shocks. 
For this, we compare the radiative cooling time, $t_{\rm cool}$, of 
relativistic electrons in the shock magnetic fields to the 
microscopic dynamical time of collisionless relativistic shocks --- 
the inverse plasma frequency of protons, $\omega_{pp}^{-1}$. 
We obtain that for $t_{\rm cool}\omega_{pp}^{-1}\la$~few hundred,
the electrons cool efficiently at or near the shock jump
and are capable of emitiing away a large fraction of the shock
energy. Such shocks are well-resolved in
existing PIC simulations; therefore, the microscopic structure 
can be studied in detail. Since most of the emission in such shocks would be coming from the vicinity of the shock, the spectral power 
of the emitted radiation can 
be directly obtained from finite-length simulations and compared with observational data.
Such radiative shocks correspond to the internal baryon-dominated 
GRB shocks for the conventional range of ejecta parameters.
Fermi acceleration of electrons in such shocks is limited by electron cooling, hence the emitted spectrum should be lacking a non-thermal tail, whereas its peak likely falls in the multi-MeV range.
Incidentally, the conditions in internal shocks are almost identical to
those in laser-produced plasmas; thus, such GRB-like plasmas 
can be created and studied in laboratory experiments using the presently 
available Petawatt-scale laser facilities.
An analysis of the external shocks shows that only the highly 
relativistic shocks, corresponding to the extremely early afterglow phase,
can have efficient electron cooling in the shock transition.  
We emphasize the importance of radiative PIC simulations for further studies. 
\end{abstract}

%\keywords{gamma rays: bursts --- shock waves --- magnetic fields}

\pacs{98.70.Rz, 52.72.+v, 52.35.-g, 95.30.Qd, 52.65.Rr}
\submitto{Astrophys. J.}
%\maketitle

\section{Introduction}

It has been shown in recent years that collisionless relativistic 
shocks are mediated by the Weibel instability --- a current filamentation 
instability that produces strong, sub-equipartition magnetic fields at the 
shock front \citep{ML99}. This is an attractive model 
for gamma-ray bursts (GRBs),
because it puts a synchrotron shock model on a firm physical ground.
However, it has been thought that the generated magnetic fields should
occupy an extremely small volume of the shocked region, hence the radiation emitted by 
the relativistic electrons in these fields should be extremely weak and 
astrophysically unimportant. Conventionally, one assumes that the strong
Weibel fields live on scales of tens of plasma skin depths 
(tens of $c/\omega_{pp}$) behind the shock, which ranges from
only few centimeters in internal shocks to about $10^8$~cm or so
in external shocks; both are some ten orders of magnitude smaller 
than the typical size of the shocked region. Therefore, unless the synchrotron 
cooling length, $t_{\rm cool}c$, is comparably small, the radiative 
efficiency of such shocks should indeed be very small. This further raises a concern 
that a number of numerical particle-in-cell (PIC) simulations attempting 
to model GRB shocks\footnote{These 2D simulations 
studying shocks in electron-position plasma are extending to at most ten  
thousand skin depths, i.e., of order tens of meters (in the shock co-moving 
frame) for internal shocks.} \citep{Spit08,CSA08,KKSW08}
probe spatial scales that are too small, and, hence,  cannot be used to deduce 
the radiation spectrum of GRB emission. This also can cast some 
doubt on jitter radiation as a viable explanation of spectral properties
of prompt GRBs \citep{M00,M06,Hededal05}. Recently, long duration shock simulations \citep{KKSW08} and analytical works \citep{KKW07,CMN08, SG07,GMf07, Milos+08} have suggested that appreciable magnetic fields could still survive in the downstream region due to the effects of accelerated particles or the field amplification by upstream turbulence. Regardless of the ultimate resolution of the magnetic decay question, there exists another physical regime where most of the radiation is produced near the shock, rather than in its downstream. Simulations of such shocks then capture most of the relevant physics without the need to simulate the far downstream region. In this paper, we explore the relevance of such shocks to GRB scenarios. 
 
We use the results of recent numerical simulations of 
relativistic collisionless shocks in electron-ion \citep{Spit08} and electron-positron \citep{CSA08,KKSW08} plasmas to estimate and 
compare the electron cooling and plasma time-scales in the shock transition region. 
We then deduce the parameter range of internal (both baryon-dominated 
and electron-positron pair-dominated) 
and external shocks, for which the shocks can be considered as 
radiatively efficient, i.e., the radiated energy in the shock transition is of order the electron 
energy $\gamma_e m_e c^2$, where $\gamma_e$ is an average 
Lorentz factor of the bulk electrons. We find that internal 
shocks are in this radiative regime for a reasonable range of model
parameters, $\Gamma\sim200-300$ and $\Gamma_i\ga2.5$, where $\Gamma$ and 
$\Gamma_i$ are, respectively, the bulk Lorentz factor of the ejecta and the 
Lorentz factor of an internal shock inside the ejecta, measured 
in the center of mass frame of the colliding shells. 
In contrast, external shocks may be in the radiative regime only 
at very early times, less than 100 seconds after the explosion.
We also argue that the physical conditions in internal shocks are very close 
to those that can be achieved in laser-plasma interaction experiments.
Therefore, the physics of the Weibel instability and jitter radiation can 
be studied experimentally at laser-plasma facilities, such as {\it Omega EP}, 
{\it NIF} and others. In particular, the {\em Hercules} experiment has been proposed and is now under development at the University of Michigan (\citealp{HerculesKU06,HerculesKU07,GRB+Hercules08}; for technical details, see \citealp{Hercules08}) to create and diagnose the Weibel instability and turbulence in the laboratory high-energy density plasmas, as a part of the Laboratory Astrophysics and High-Energy Density Physics programs.

\section{Dimensionless electron cooling time}

The electron synchrotron cooling time is obtained from the relation
that the emitted energy is of order the initial electron's energy:
\beq
P_{\rm syn}t_{\rm cool}=(4/3)\sigma_T c \gamma_e^2(B^2/8\pi) t_{\rm cool}
\sim \gamma_e m_e c^2.
\eeq
Here, $\sigma_T$ is the Thompson cross-section, $\gamma_e$ is the 
Lorentz factor of the emitting electron, and $P_{\rm syn}$ is the 
synchrotron emission power per electron.
This gives
\beq
t_{\rm cool}=\frac{6\pi m_e c}{\sigma_T\gamma_e B^2}.
\eeq
Note that this expression holds for jitter radiation as well, because
the total emitted power by an electron in the jitter regime is identical
to that of synchrotron \citep{M00}.

We will compare $t_{\rm cool}$ to the characteristic time of the plasma 
processes at a shock, the inverse {\em relativistic} plasma frequency 
of the protons, $\omega_{pp}^{-1}$, which is
\beq
\omega_{pp}=\left(\frac{4\pi e^2 n'}{\Gamma_i m_p}\right)^{1/2},
\eeq
where $n'$ is the particle density behind the internal shock measured 
in the downstream frame. To within a factor of two, the so-defined
$\omega_{pp}$ corresponds to the nonrelativistic upstream plasma
frequency, because $n'=4 n \Gamma_i$, where $n$ is the particle
density of the unshocked ejecta in its own co-moving frame. 
It is natural to introduce the dimensionless cooling time
\beq
T_{\rm cool}=t_{\rm cool} \omega_{pp}.
\eeq

In our analysis we will use the results of electron-ion \citep{Spit08} and electron-positron  \citep{CSA08} shock PIC simulations. Apparently,
the evolution of currents and magnetic fields in the vicinity of a shock 
in electron-proton and electron-positron plasmas does not differ 
substantially, because in the electron-proton case the electrons 
in the downstream carry about $30\%-50\%$ of the proton energy and, hence, 
their effective mass is comparable to the proton mass.\footnote{
One has to be careful here because the parallel and perpendicular 
relativistic masses are not identical, 
$m_{e,\perp}\sim\gamma_e m_e\sim\Gamma_i m_p$ and 
$m_{e,\|}\sim\gamma_e^3 m_e\sim\Gamma_im_p(\Gamma_im_p/m_e)^2\gg\Gamma_im_p$.
Hence, the electrons will be effectively much heavier than the 
protons with respect to their acceleration/deceleration, 
but should behave similar to the protons 
with respect to deflection and pitch-angle scattering.}

\begin{figure}
\includegraphics[width=5.5in]{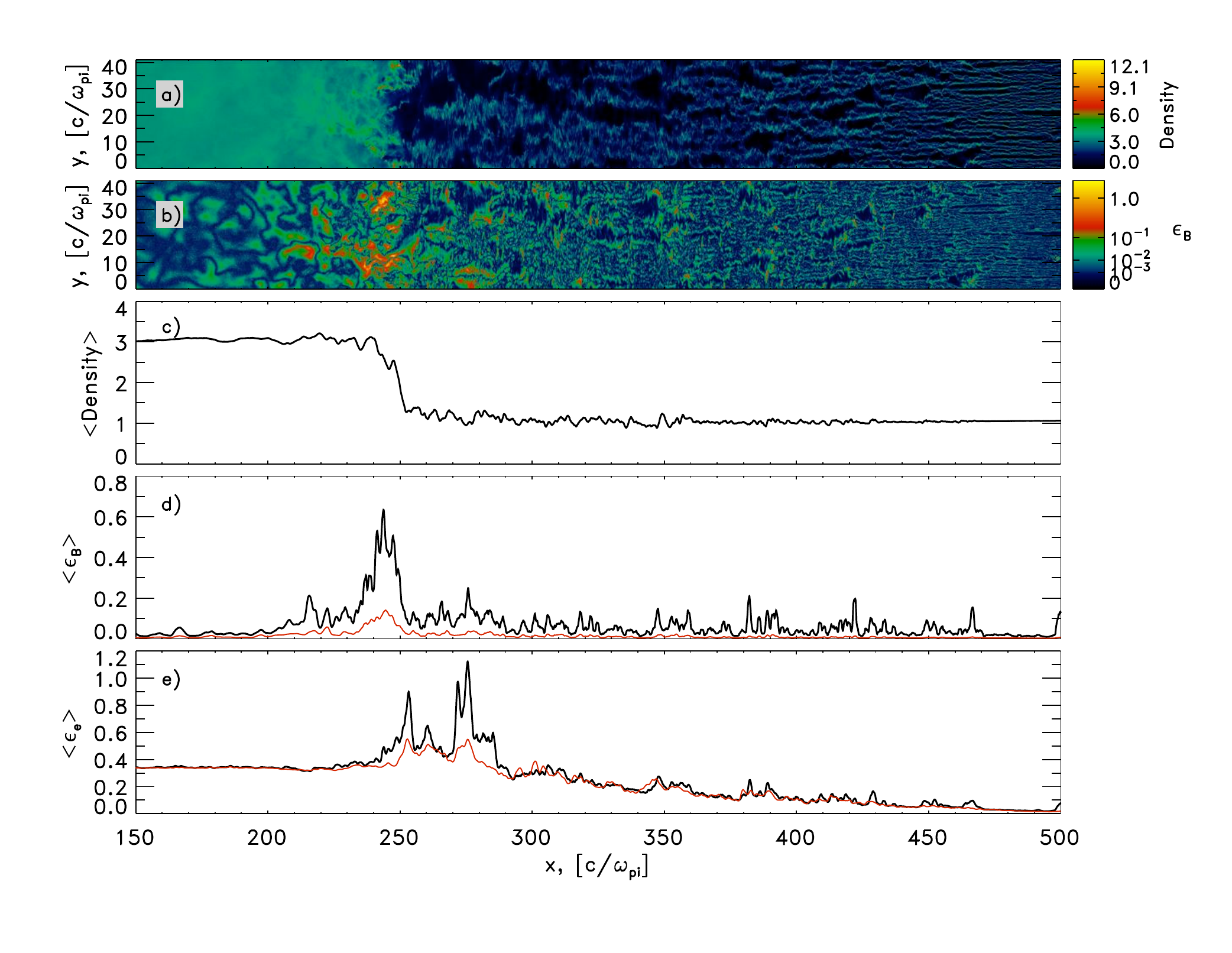}
\caption{A snapshot of a shock from PIC simulation with $\Gamma=15$ and $m_i/m_e=100$. 
(a) and (b) --- Maps of the particle density $n$ and the normalized magnetic field strength $\epsilon_B$, respectively. The incoming flow moves from right to left, while the shock propagates to the right. The simulation is performed in the downstream frame. (c) -- Density profile averaged over the transverse ($y$) direction. 
(d) and (e) --- The $y$-averaged profiles of $\epsilon_B$ and $\epsilon_e$ (red curves) and the ``emission-weighted" profiles of $\hat{\epsilon}_B$ and $\hat{\epsilon}_e$ (black curves) respectively.
\label{f0}}
\end{figure}

Figure \ref{f0} represents a snapshot of a steady state shock in the electron-ion plasma obtained from a PIC simulation with Lorentz factor $\Gamma=15$ and mass ratio $m_i/m_e=100$ \citep{Spit08}. 
This and other simulations \citep{CSA08,KKSW08} indicate that the strongest magnetic fields with 
energy density $\ga 10\%$ of the kinetic energy density occupy the region of few tens of ion plasma skin-depths 
around the main shock compression, as measured in the frame of the 
downstream fluid. Hereafter, we will use the value of $l\sim50c/\omega_{pp}$ for the transition length (for electron-positron shocks, the ion plasma frequency is replaced with the plasma frequency of pair plasma).
Since the fields are highly 
inhomogeneous, we assume a factor of two uncertainly in this number.
The shock moves at $v=c/3$ (or $c/2$ in 2D) in the downstream frame, 
hence the electron residence time in the region of high field is 
$t_{\rm res}\sim l/v\sim150\omega_{pp}^{-1}$. Taking the uncertainty in
$l$ and other parameters into account, we estimate that
$t_{\rm res}\sim(100-300)\omega_{pp}^{-1}$. We did not account here for
the clumpiness of magnetic inhomogeneities, which shortens the effective 
$t_{\rm res}$, and the electron trapping in high-field clumps,
which is increasing the effective residence time.
Interestingly, in the far
downstream region, the magnetic energy decreases with the distance $d$
from the shock as $\propto d^{-1}$ \citep{CSA08}. An electron in this region loses 
energy logarithmically slowly, $E\propto\ln(d)$. If this scaling holds 
through a large distance downstream, this implies two possibilities. 
First, if $t_{\rm cool}\la t_{\rm res}$, 
then the electrons lose their energy quickly near
the shock jump; hence the radiative efficiency of such a shock is high.
The electron cooling will substantially affect the structure of the 
downstream region, in this regime. Second, if $t_{\rm cool}\gg t_{\rm res}$, 
then the electrons lose only a small 
fraction of their energy, which makes this shock radiatively inefficient.
Therefore, we refer to a shock as a ``radiative shock'' if 
$T_{\rm cool}\la300$ and as a ``weakly radiative shock'' otherwise.
Interestingly, if $T_{\rm cool}\la10-50$, radiative cooling might be substantial
already in the upstream region (where the Weibel instability creates 
current filaments and magnetic fields), before the main shock compression. 
We expect that the very formation of such a shock may be strongly affected 
by radiative cooling of electrons. We refer to the regime of $T_{\rm cool}\la10-50$
as the ``strong cooling regime.''

\section{Shock models}
\subsection{Internal shocks in baryon-dominated ejecta} 

Internal shocks occur inside the ejected material when inhomogeneities, 
often referred to as ``shells' that have different masses 
and velocities, collide with each other relativistically.
Radiation emitted at the internal shocks is thought to be the 
prompt gamma-ray radiation of a GRB (e.g., \citealt{Mes06}).

Here we conventionally assume that a central engine produces an
ultra-relativistic wind with the kinetic luminosity, 
$L\sim10^{52}$~erg/s and the Lorentz factor $\Gamma\ga100$
\citep{MR00,RM05,PW04}.
At a radial distance, $R$, from the central engine, the 
co-moving particle density of the ejecta is 
\beq
n=\frac{L}{4\pi R^2 \Gamma^2 m_p c^3}
\simeq(1.8\times 10^{15}~{\rm cm}^{-3}) L_{52}R_{12}^{-2}\Gamma_2^{-2},
\eeq
where we use the convention that $L_{52}=L/(10^{52}~{\rm erg/s})$ 
and similarly for other quantities, whereas if no numerical subscript is 
present, then the quantity is in CGS units. The density in the 
downstream of an internal shock is 
\beq
n'=4\Gamma_i n.
\eeq
The magnetic field in the downstream and the electron bulk Lorentz 
factor are calculated as fractions $\epsilon_B$ and $\epsilon_e$ 
of the post-shock thermal energy density, 
$U=n'(\gamma_p m_pc^2+\gamma_e m_ec^2)\sim n'\Gamma_im_pc^2$
(in the last equality, we neglected the electron contribution, for
simplicity, though it may change $U$ by a factor $\epsilon_e/[1+\epsilon_e]$),
\bea
B'&=&\left(8\pi\Gamma_i m_pc^2n'\epsilon_B\right)^{1/2}
\simeq(1.6\times10^7~{\rm G})L_{52}^{1/2}\Gamma_2^{-1}
R_{12}^{-1}\epsilon_B^{1/2}.
\\
\gamma_e&=&(m_p/m_e)\Gamma_i\epsilon_e\simeq1.8\times10^3\Gamma_i\epsilon_e.
\eea
For the parameters $\epsilon_e$ and $\epsilon_B$ one has to use the
emission-weighed quantities, rather than the conventionally used 
volume-averaged
ones, because fields are highly inhomogeneous in the shock region.
This effect is particularly important for $\epsilon_B$ and less so for
$\epsilon_e$ because high-energy electrons are distributed more or less 
uniformly. PIC simulations indicate that $\epsilon_e\sim0.5$ 
(electrons carry 50\% of the proton energy) and that $\epsilon_B$ reaches 
unity locally \cite{Spit08,CSA08}. High energy radiation from electrons is not included in the present code. In order to calculate the emission-weighted $\hat{\epsilon}_e$ and $\hat{\epsilon}_B$ (black curves in panels (d) and (e) in figure \ref{f0}, hereafter denoted with a ``hat" symbol) we took into account that radiative losses are proportional to 
$B^2\gamma_e^2$. Since the electron cooling occurs predominantly in the high-field 
regions, a larger effective $\hat{\epsilon}_B$ and $\hat{\epsilon}_e$ would be deduced from observations. 
Taking into account the factor of 8 difference in the definition
of $\epsilon_B$ here and in \citealt{Spit08,CSA08}, we estimate the 
emission-weighted value as $\hat{\epsilon}_B\sim1/8\sim0.1$.

Finally, the dimensionless cooling time in 
baryon-dominated internal shocks becomes
\beq
T_{\rm cool}^{(e^-p)}\simeq170 L_{52}^{-1/2}\Gamma_2\Gamma_i^{-3}
R_{12}\hat{\epsilon}_B^{-1}\hat{\epsilon}_e^{-1}.
\label{Tc-ep}
\eeq

We assume that the shells have different masses but carry similar 
linear momenta, $p\sim \Gamma_s m_sc\sim\Gamma_r m_rc$. That is, 
in this scenario, a constant driving force is assumed to continuously
eject material while the central engine is active.
Thus, the variation in the Lorentz 
factor of the shells is simply reflecting the variation of the  
density of the ejecta (i.e., that of the shell masses). An attractive 
feature of this model is that the total energy of the ejecta is uniformly 
distributed among individual shells, which nicely fits into the relativistic
wind picture, unlike the model with shells of equal mass, in which the 
fastest shell essentially dominates the energetics of the outflow.  
In the collision of two shells, two shock waves are formed, which 
propagate through each shell \citep{KPS97}. The center of mass Lorentz factor 
of the two shells is 
\beq
\Gamma_c=\frac{m_s\Gamma_s+m_r\Gamma_r}%
{\left[m_s^2+m_r^2+m_s m_r(\Gamma_s/\Gamma_r+\Gamma_r/\Gamma_s)\right]^{1/2}},
\eeq
so $\Gamma_c\sim\sqrt{2}\Gamma_s$ for $\Gamma_s m_s\simeq\Gamma_r m_r$
and $\Gamma_r\gg\Gamma_s$.
The thermal energy density in the
downstream of each shock is related to the Lorentz factor of
the shells in their center of mass frame:
\beq
(\Gamma_i)_{s,r}\simeq(1/2)(\Gamma_{s,r}/\Gamma_c+\Gamma_c/\Gamma_{s,r}),
\eeq
so that the Lorentz factor of an internal shock can be as high as 
$\Gamma_i\sim 2^{-3/2}\Gamma_r/\Gamma_s$, which is at least a few if
the Lorentz factors of the ``slow'' and ``rapid'' shells are,
say, $\Gamma_s\sim100$ and $\Gamma_r\sim1000$, respectively, that
is $m_s/m_r\sim10$.

We assume that the central engine is an accretion disk around a 
solar mass black hole. Thus, the flow forms at a few to ten Schwarzschild 
radii. This sets the variability time scale, $t_v\sim10^{-4}$~s. 
Present observations confirm the variability on a millisecond scale,
but are not capable of resolving sub-millisecond time scales. 
In this model, the relativistic shells are ejected from the radius 
$R_0\sim ct_v$ and their initial separation is of order $R_0$ as well. 
A collision of the two shells will occur at the radius
\beq
R_i\simeq 2R_0\left(\Gamma_s^{-2}-\Gamma_r^{-2}\right)^{-1}
\sim 2\Gamma_s^2ct_v\sim (6\times10^{10}~{\rm cm})\Gamma_2^2 t_{v,-4}.
\eeq
One should keep in mind that collisions of the shells can occur 
at substantially smaller radii, if the wind modulation occurs
due to instabilities in the outflow (e.g., a jet), at radii
$R\gg R_0$ where the flow is already ultra-relativistic \citep{RM05}. 

Due to the presence of electrons in the baryonic outflow, the 
co-moving optical depth due to Thompson scattering, 
$\tau_b\sim n\sigma_T (R/\Gamma)$, approaches unity at the radius
\beq
R_{\rm ph}=\Gamma/(n\sigma_T)
\simeq(1.2\times10^{13}~{\rm cm})L_{52}\Gamma_2^{-3},
\eeq
where $\sigma_T$ is the Thompson cross-section and $R/\Gamma$ is 
the co-moving length. For a radiative luminosity of a GRB of 
$L_\gamma\sim0.1 L$ (we assumed a 10\% radiation efficiency), the
electron-positron pair opacity is obtained from the balance
between the rates of annihilation and of pair production. The latter
is the photon density above the electron rest mass over the
co-moving dynamical time \citep{PW04}. Taking into account that 
only the photons with energies above $m_e c^2$ can produce pairs, 
the pair-producing luminosity is 
$L_\gamma^>\simeq (\hbar\nu_{\rm syn}/m_ec^2)^{\beta}L_\gamma$ 
if $\hbar\nu_{\rm syn}<m_ec^2$ and $L_\gamma^>=L_\gamma$ otherwise, where
$\nu_{\rm syn}=\gamma_e^2(eB'/m_ec)$ is the co-moving synchrotron frequency,
$\beta\sim0.5$ is the high-energy spectral slope 
$\nu F_\nu\propto\nu^{-\beta}$. Finally, we obtain the pair optical
depth to be equal to the square root of the co-moving compactness,
$\tau_\pm\sim l'^{1/2}
\sim\left(L_\gamma^>\sigma_T/4\pi R\Gamma^3m_pc^3\right)^{1/2}$.
The radius of the pair photosphere, $\tau_\pm\sim1$, is 
\beq
R_{\rm pair}=L_\gamma^>\sigma_T/4\pi \Gamma^3m_pc^3
\simeq(2.2\times10^3~{\rm cm})L_{\gamma,51}\Gamma_2^{-3}.
\eeq
In the last expression we omitted the spectral correction, for simplicity.
Note that non-thermal spectra can be seen even for relatively
large $\tau_\pm$ of few tens \citep{PW04}. Since the
optical depth increases with decreasing radius as
$\tau_\pm\propto l'^{1/2}\propto R^{-1/2}$, one can observe non-thermal 
spectra produced at internal shocks at radii as small as 
$\sim10^{-3}R_{\rm pair}$ or so. Note also that the position of 
the pair photosphere depends on (unknown) radiative efficiency.

Figure \ref{f1} shows the regions of strong and weak radiative cooling 
versus $R$ and $\Gamma$ for a reasonable and a rather extreme cases of $\Gamma_i=4$ and $\Gamma_i=10$. One can see that internal shocks in the strong cooling regime , $T_{\rm cool}\la10$, 
can occur for $\Gamma\la10^{2.1}\sim125$ and at
$R\la{\rm few}\times(10^{10}-10^{11})$, well below the photosphere at
such low $\Gamma$'s; hence, radiation from
such shocks is hardly observable. Strongly radiative shocks 
$T_{\rm cool}\sim100-300$ can occur above the baryonic photosphere
for reasonable values of $\Gamma\la200-300$ and $\Gamma_i\ga2.5$; 
hence they are likely observable. At the photosphere, $R_{\rm ph}=R_i$, 
we have
\bea
R_{i,\rm ph}&\simeq&(4.9\times10^{11}~{\rm cm})L_{52}^{2/5}t_{v,-4}^{3/5},
\\
\Gamma(R_{i,\rm ph})&\simeq&290L_{52}^{1/5}t_{v,-4}^{-1/5},
\eea
which depend weakly on the central engine parameters.
Excluding $R$ and $\Gamma$ in Eq. (\ref{Tc-ep}), we obtain
\beq
T_{\rm cool}^{(e^-p)}(R_{i,\rm ph})\simeq 250L_{52}^{1/10}t_{v,-4}^{2/5}
\Gamma_i^{-3}\hat{\epsilon}_B^{-1}\hat{\epsilon}_e^{-1},
\eeq
We plot the the regions of strong and weak cooling versus $R$ and $\Gamma_i$ in the left panel of figure \ref{f2} and $T_{\rm cool}$ vs. $\Gamma_i$ in right panel. We conclude
that internal shocks with $\Gamma_i$ as low as 2.5 can be 
radiatively efficient.

The peak of the synchrotron radiation in the observers frame is
\beq
\nu_{syn}=(eB/m_ec)\gamma_e^2\Gamma
\simeq400~{\rm MeV}~ L_{52}^{1/2}R_{12}^{-1}\Gamma_i^{3}\hat{\epsilon}_B^{1/2}\hat{\epsilon}_e^{2}.
\eeq
If radiation is emitted in the jitter regime, the spectral peak is expected to be a bit higher:
$\nu_j\simeq\nu_{syn}\sqrt{(m_i/m_e)\hat{\epsilon}_B}\sim10\nu_{syn}$ \citep{M00,M+07}.
Thus, for the assumed parameters the emission is expected to be peaked at few tens to few hundreds MeV, if the shock is in the radiatively efficient regime with $\Gamma\sim300$ and $R\sim10^{12}~{\rm cm}$.

\subsection{Internal shocks in pair-dominated outflows} 

This case can be readily obtained from the baryon-dominated case by replacing 
$m_p$ with $m_e$. We obtain
\beq
T_{\rm cool}^{(e^\pm)}\simeq5.9\times10^8 L_{52}^{-1/2}\Gamma_2\Gamma_i^{-3}
R_{12}\hat{\epsilon}_B^{-1}\hat{\epsilon}_e^{-1}.
\label{Tc-ee}
\eeq
Thus, internal shocks in the $e^\pm$-dominated ejecta 
have a very long radiative cooling time and, hence, are radiatively
inefficient in our language.
The peak of the synchrotron radiation in the observers frame is
\beq
\nu_{syn}
\simeq0.12~{\rm keV}~ L_{52}^{1/2}R_{12}^{-1}\Gamma_i^{3}\hat{\epsilon}_B^{1/2}\hat{\epsilon}_e^{2}.
\eeq

\subsection{External shocks of afterglows} 

In the post-shock region of an external shock propagating into an
external medium of density $n_{\rm ext}$, we have
\bea
n'&=&4\Gamma n_{\rm ext}=(400~{\rm cm}^{-3})\Gamma_2 n_{\rm ext},
\\
\gamma_e&=&(m_p/m_e)\Gamma\hat{\epsilon}_e=1.8\times10^{5}\Gamma_2\hat{\epsilon}_e,
\\
B'&=&(8\pi\Gamma m_pc^2n'\hat{\epsilon}_B)^{1/2}\simeq(39~{\rm G})
\Gamma_2n_{\rm ext}^{1/2}\hat{\epsilon}_B^{1/2},
\eea
Using these quantities, we estimate the cooling time as
\beq
T_{\rm cool}^{\rm AG}\simeq
7.3\times10^{3}\Gamma_2^{-3}n_{\rm ext}^{-1/2}\hat{\epsilon}_B^{-1}\hat{\epsilon}_e^{-1}.
\label{Tc-AG}
\eeq
This relation is plotted in figure \ref{f3} for a very dense external medium,
$n_{\rm ext}\sim100~{\rm cm}^{-3}$. Even for such an extreme case, 
the external shock is radiative only for very large Lorentz factors 
$\Gamma\ga500$. It is an order of magnitude larger for a more 
conventional value of $n_{\rm ext}\sim1~{\rm cm}^{-3}$. 

We now consider two models of the external medium density profile, namely
the constant density interstellar medium (ISM) and the wind outflow models.

In the {\it ISM model}, the Lorentz factor of a blast wave depends 
on the observed time as \citep{GPS99}
\beq
\Gamma\simeq350 E_{53}^{1/8} n_{\rm ISM}^{-1/8}(1+z)^{3/8}t^{-3/8},
\eeq
where $E\simeq Lt_{\rm GRB}$ is the isotropic energy equivalent 
of the blast wave, $t_{\rm GRB}$ is the duration of a GRB 
and $n_{\rm ext}=n_{\rm ISM}=$~const 
is the external medium density.
In this model, the cooling time is 
\beq
T_{\rm cool}^{\rm AG, ISM}\simeq
180 E_{53}^{-3/8}n_{\rm ISM}^{-1/8}(1+z)^{-9/8}
\hat{\epsilon}_B^{-1}\hat{\epsilon}_e^{-1}t^{9/8}.
\label{Tc-AGISM}
\eeq

In the {\it Wind model}, the blast wave is propagating in the wind environment 
with the density decreasing with distance as $n\propto R^{-2}$; the Lorentz
factor  of the blast wave and wind density are \citep{CL00}
\bea
\Gamma&\simeq&270 E_{53}^{1/2} A_*^{-1/4}(1+z)^{1/4}t^{-1/4},
\\
n_{\rm ext}&\simeq&(1.1\times10^5~{\rm cm}^{-3}) 
E_{53}^{-1} A_*^{2}(1+z)t^{-1},
\eea
where  $A_*=[\dot M_W/(10^{-5}M_{\sun}\textrm{
yr}^{-1})] / [V_W/(10^3\textrm{ km s}^{-1})]$ is the wind parameter, 
$\dot M_W$ is the mass loss rate, $V_W$ is the wind velocity. 
The cooling time is
\beq
T_{\rm cool}^{\rm AG, W}\simeq
1.2E_{53}^{-1}A_*^{-1/4}(1+z)^{-5/4}\hat{\epsilon}_B^{-1}\hat{\epsilon}_e^{-1}t^{5/4}.
\label{Tc-AGW}
\eeq

In figure \ref{f4}, we plot $T_{\rm cool}$ in the external shock 
versus time after the burst for both models and for two sets of afterglow 
parameters. We see that, except for the very early times, in both models, 
the emission from external shocks should be coming from far downstream, 
not from the main shock compression region.
However, in the Wind model, the external shock can be radiative
up to $\sim100$~s after the burst, whereas for the ISM model, the radiative
shock regime can occur only at times earlier than a second after the explosion.
Since the afterglow usually sets in at least several 
tens of seconds after the explosion,
we conclude that only very early afterglow emission can, in principle, 
come from radiative shocks and only in the Wind model.

\section{Discussion}

We compared the electron cooling time to the dynamical microscopic
time (the plasma time, $\omega_p^{-1}$) for the internal shocks
in both electron-positron and baryon-dominated relativistic GRB 
outflows and for the external afterglow shock. We used the most
recent PIC simulations to obtain $\epsilon_e$,\ $\epsilon_B$ and the
size of the region with strong magnetic field. We evaluated and used 
in our analysis the emission-weighted $\hat{\epsilon}_B$, instead of the 
conventional volume-averaged quantity (for $\epsilon_e$ both emission- 
and volume-averaged quantities are expected to be similar). 
We then evaluated the 
residence time of an electron in the high-field region as the size of
this region over the flow speed. Thus, we did not account for
clumpiness and less-then-unity filling factor of magnetic inhomogeneities 
(which is decreasing the effective residence time) and the
electron trapping in high-field clumps (which is increasing the
effective residence time). Using PIC simulations, one can study 
the radiative cooling effects much more accurately. 

We obtained that if
$T_{\rm cool}=t_{\rm cool}\omega_p<(100-300)$, 
an electron has enough time to radiate away energy 
comparable to its initial energy, $\sim\gamma_em_ec^2$. 
If $(10-50)<T_{\rm cool}<(100-300)$, much radiation is emitted from the
strong field region at the shock jump. Therefore, we
call this regime the ``radiative shock'' regime.
If $T_{\rm cool}\la(10-50)$, the electron cooling is extremely fast and 
strong radiative losses are expected  in the 
upstream region as well; hence we refer
to it as the ``strong cooling regime''.
Note that in both these cases, Fermi acceleration of electrons is hardly possible because the electrons lose their energy every time they cross the shock. Therefore, we expect that the electron distribution will not be a power-law and, hence, a hard non-thermal tail in the radiation spectrum is hardly produced.
If $T_{\rm cool}\gg300$,
radiation from the shock (if any) will be coming from the far downstream
region. This region has not been fully analyzed in simulations, so it is 
too early to draw any firm conclusions about this regime. However, 
if we extrapolate the decay rate $\epsilon_B\propto 1/d$ ($d$ is the 
distance from the shock front) as seen in PIC $e^\pm$ simulations, then 
the shocks in this regime should be very weakly radiative. 

The obtained  results are as follows. For internal shocks in 
$e^\pm$-pair-dominated
outflows, the cooling time exceeds the microphysics time by many
orders of magnitude, $T_{\rm cool}\gg300$ for the entire range
of reasonable shock and outflow parameters. Therefore, emission from
such pair-dominated shocks will have to be produced (if at all) in
the far downstream region, not at the shock front. 

Somewhat similar conclusions
follow for the external afterglow shocks in the electron-proton plasma.
For most of the blast wave conditions, these shocks are also in the
weakly radiative regime. The observed afterglow emission has to be
produced in the large volume of the downstream region. The emission from
the high-field region at the shock jump is strong for ultra-relativistic 
shocks with $\Gamma\ga350$ in a high-density external medium, 
$n_{\rm ext}\sim100~{\rm cm}^{-3}$, and for even higher $\Gamma$'s 
at lower densities. Figure \ref{f3} shows $T_{\rm cool}$
as a function of $\Gamma$. Dark and light blue regions indicate 
the range of $\Gamma$ for which an external shock is in the 
radiative regime. 

The cooling time for the ISM and Wind models of the ambient density profile
as a function of time after the burst is shown in figure \ref{f4}.  
We obtained that, except for the very early times, the shocks (in both models) 
are characterized by very large $T_{\rm cool}$, indicating that 
the observed emission should be produced in the far downstream region.
In the ISM model, the shock can be radiative only at the very early times,
earlier than a second after the burst. 
In contrast, the shock in the Wind model can be in the radiative regime 
until about a hundred seconds after the explosion. Usually, the afterglow 
sets in at least a few tens of seconds after the explosion.
Therefore, there is a chance that the very early afterglow emission from
a shock in the Wind medium is produced at the shock front itself.
We speculate that one can expect a break in the afterglow (steepening 
of the light-curve) at about 10 to 100 seconds, indicating the transition 
from the radiatively efficient to the radiatively inefficient shock.
It is likely, though, that this early afterglow emission will be
swamped in a brighter prompt or high-latitude emission. 
It is far too early to make quantitative predictions, however, because
radiative cooling can change the shock formation and evolution on a 
microscopic scale. A further study requires a radiative PIC simulation. 
It is quite possible that magnetic fields may be produced by the relativistic Richtmeyer-Meshkov instability (the vorticity-generating fluid instability, \citealp{SG07,GMf07,Milos+08}) in the afterglow phase. The dependence $\epsilon_B\propto\Gamma^{-1}$ can be a benchmark of the model.

\begin{figure}
\includegraphics[width=2.5in]{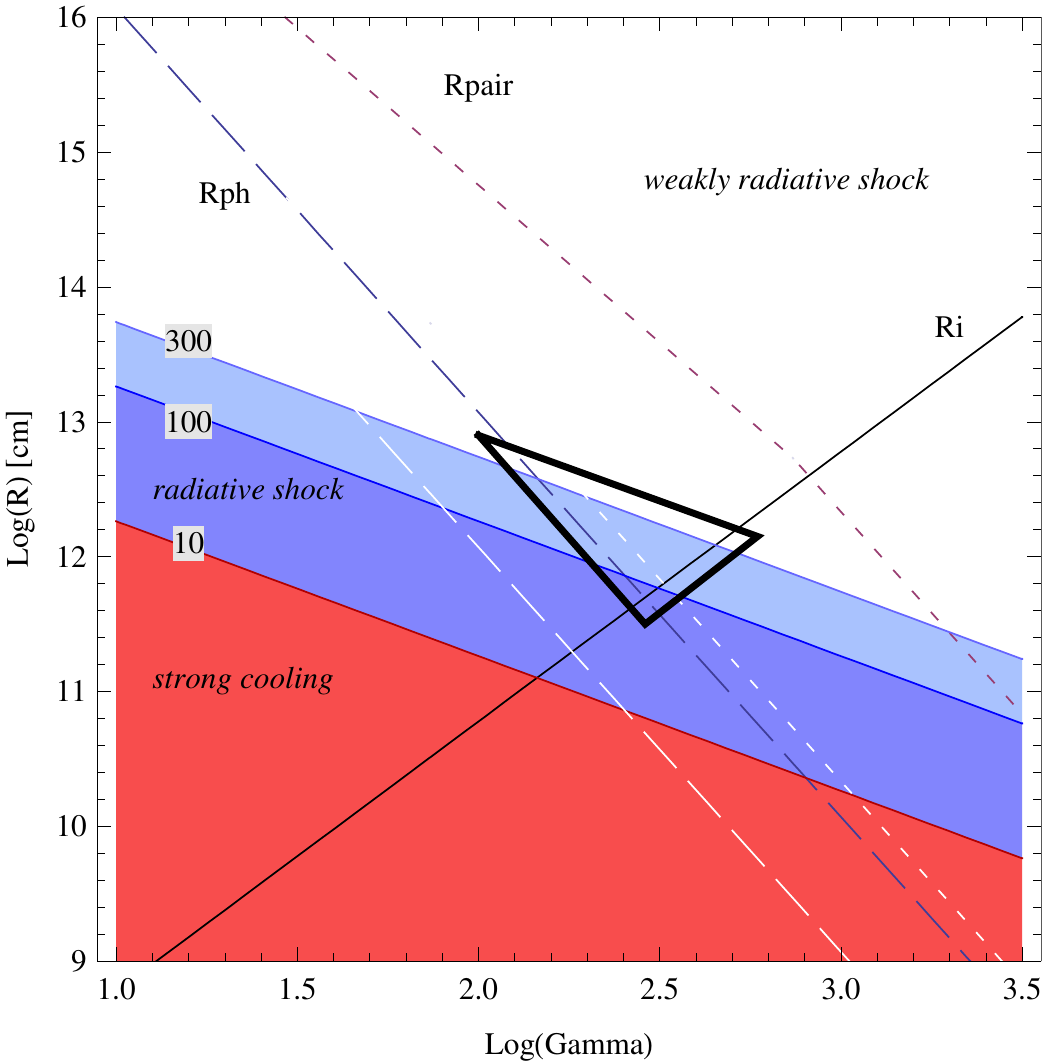}\hfill
\includegraphics[width=2.5in]{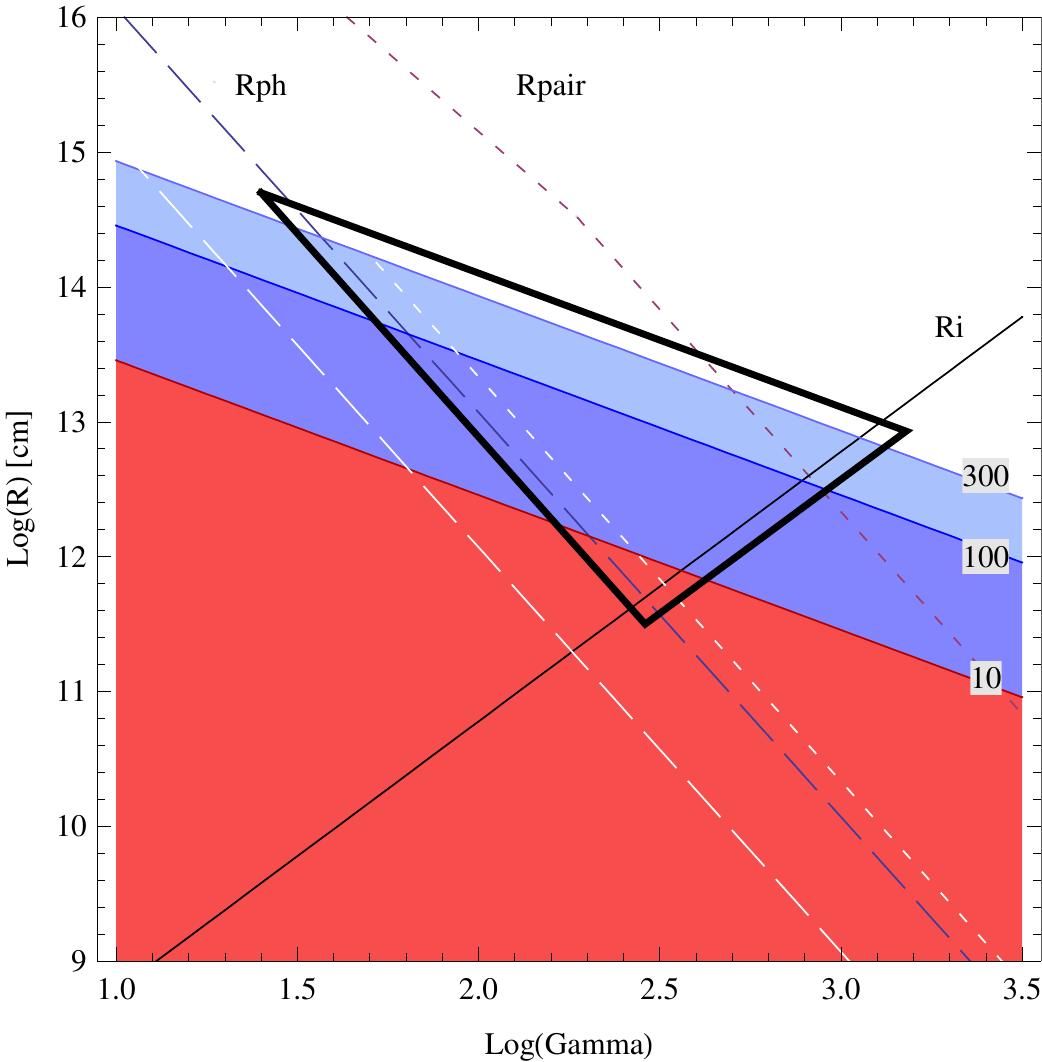}
\caption{Contours of $T_{\rm cool}$ vs. $\Gamma$ for $\Gamma_i=4$ (left panel) and a somewhat unrealistically high
$\Gamma_i=10$ (right panel) for the internal shocks in 
baryon-dominated outflows for $T_{\rm cool}=10,\ 100,\ 300$. 
Red filled regions correspond to
$T_{\rm cool}<10$ (the ``strong cooling'' regime), dark and light blue 
regions correspond to $10<T_{\rm cool}<100$ and $100<T_{\rm cool}<300$,
respectively (the ``radiative shock'' regime), and the white
region corresponds to the ``weakly radiative shock'' ($T_{\rm cool}>300$).
The black triangle outlines the range of parameters where radiative shocks occur and may be observable.
In all cases, $L_{52}=1,\ L_\gamma=0.1L,\ t_{v,-4}=1$.
The radii, $R_i$, $R_{\rm ph}$ and $R_{\rm pair}$ are plotted for reference.
Internal shocks can occur above $R_i$ line; optical depths, $\tau_b$ 
and $\tau_\pm$, are less than unity above lines
$R_{\rm ph}$ and $R_{\rm pair}$, respectively.   
White lines in the left panel denote radii at which the optical depths $\tau_b$ 
and $\tau_\pm$ are equal to 10 (dashed and dotted lines respectively). It is expected that some non-thermal spectral signatures of radiation emitted from not so large optical depths of $\tau\lesssim10$ may still be observable in addition to the strong thermal component.   
\label{f1}}
\end{figure}

\begin{figure}
\includegraphics[width=2.5in]{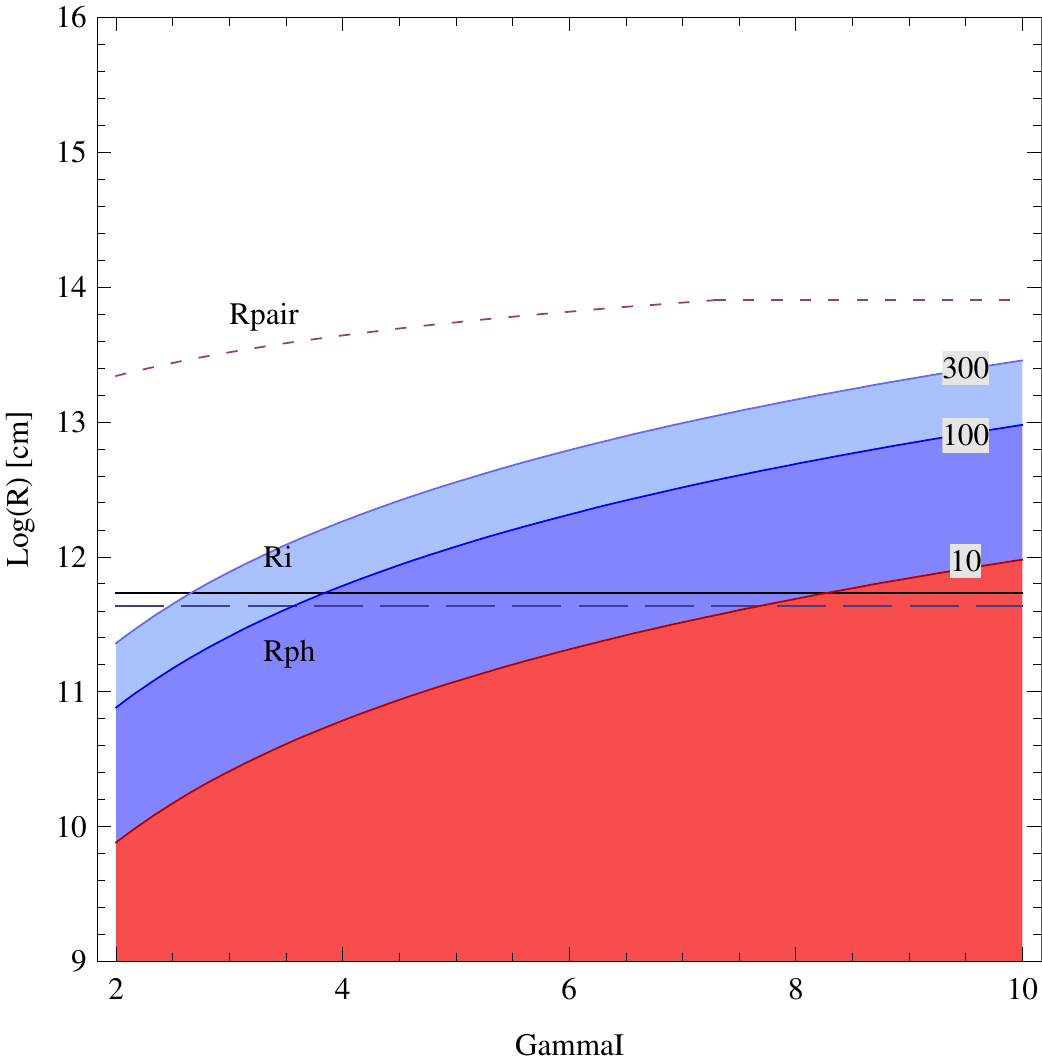}\hfill
\includegraphics[width=2.5in]{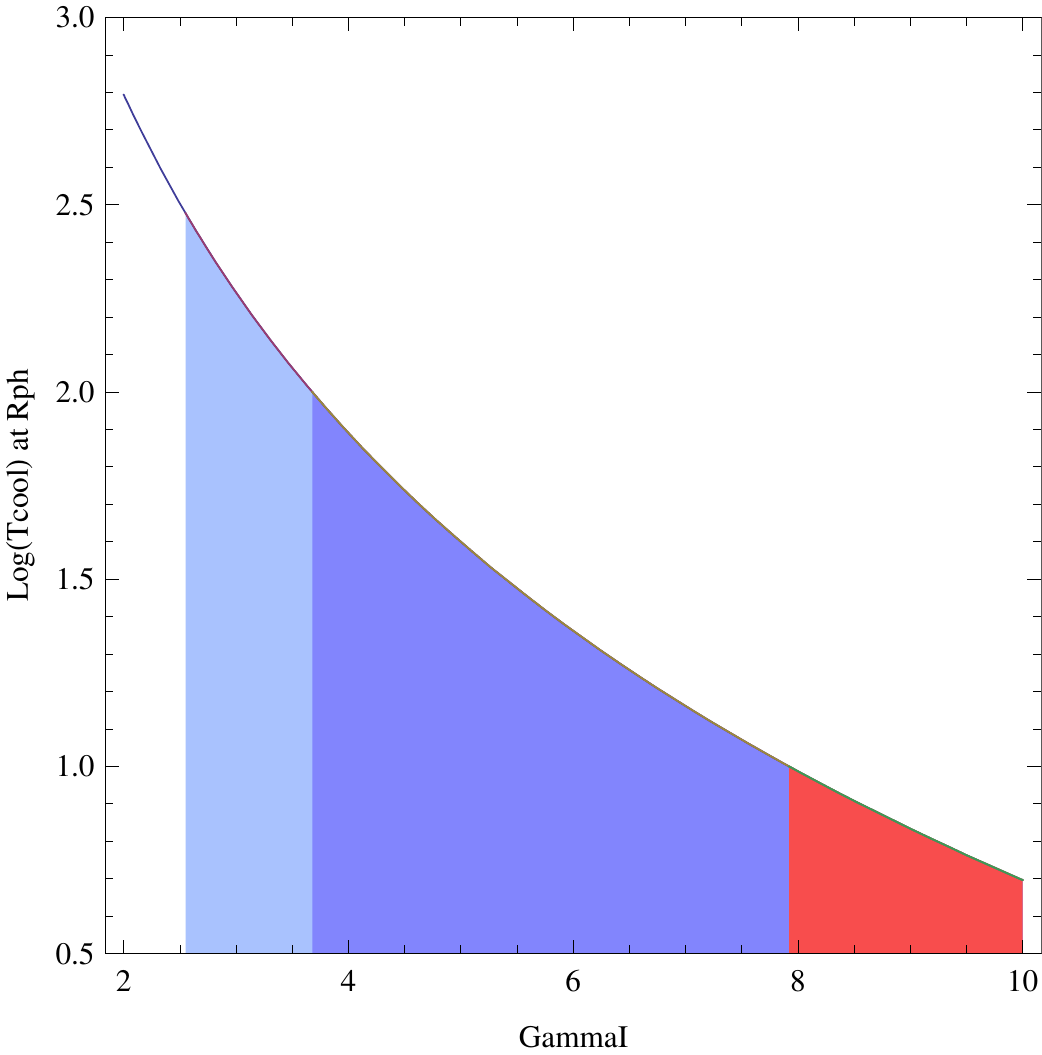}
\caption{(left panel) Contours of $T_{\rm cool}$ vs. the Lorentz factor of internal shocks, $\Gamma_i$, for the internal shocks in the $\Gamma=300$ ejecta.  (right panel) Dimensionless cooling time, $T_{\rm cool}$, in internal shocks at the baryonic photosphere vs. $\Gamma_i$. 
Color coding and the outflow parameters are the same as in figure \ref{f1}.
\label{f2}}
\end{figure}

\begin{figure}
\includegraphics[width=2.5in]{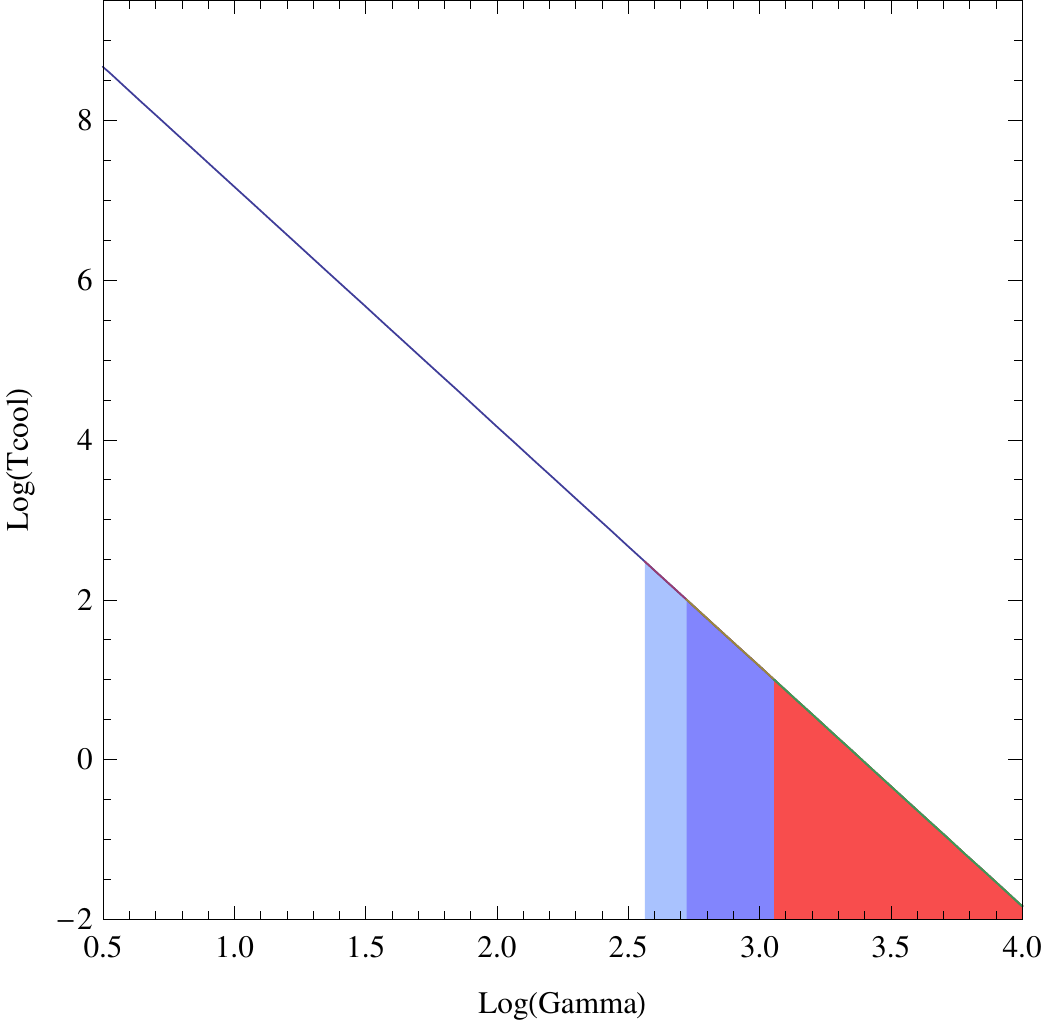}
\caption{Cooling time in afterglows vs. the outflow Lorentz factor, for
$n_{\rm ext}=100~{\rm cm}^{-3}$. Color coding and the outflow parameters are 
the same as in figure \ref{f1}.
\label{f3}}
\end{figure}

\begin{figure}
\includegraphics[width=2.5in]{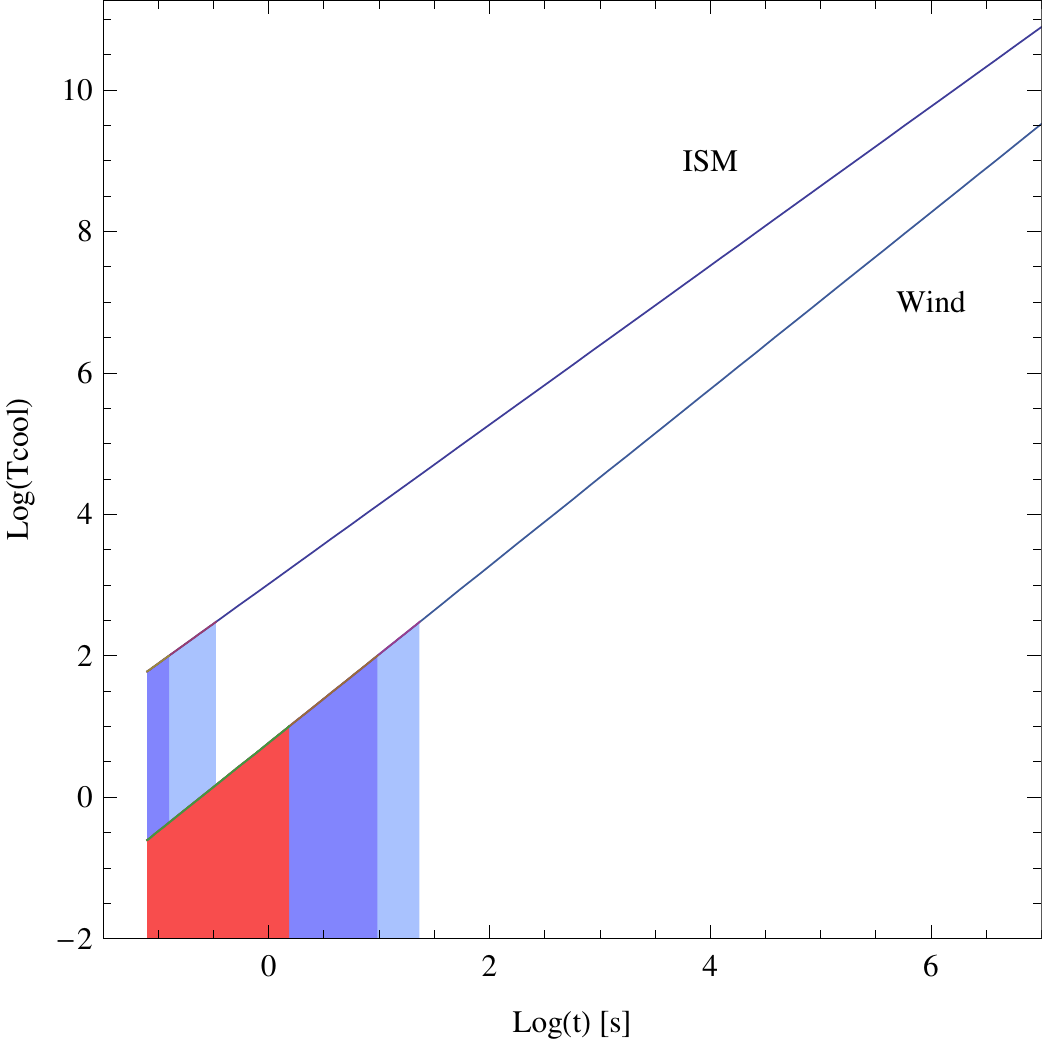}\hfill
\includegraphics[width=2.5in] {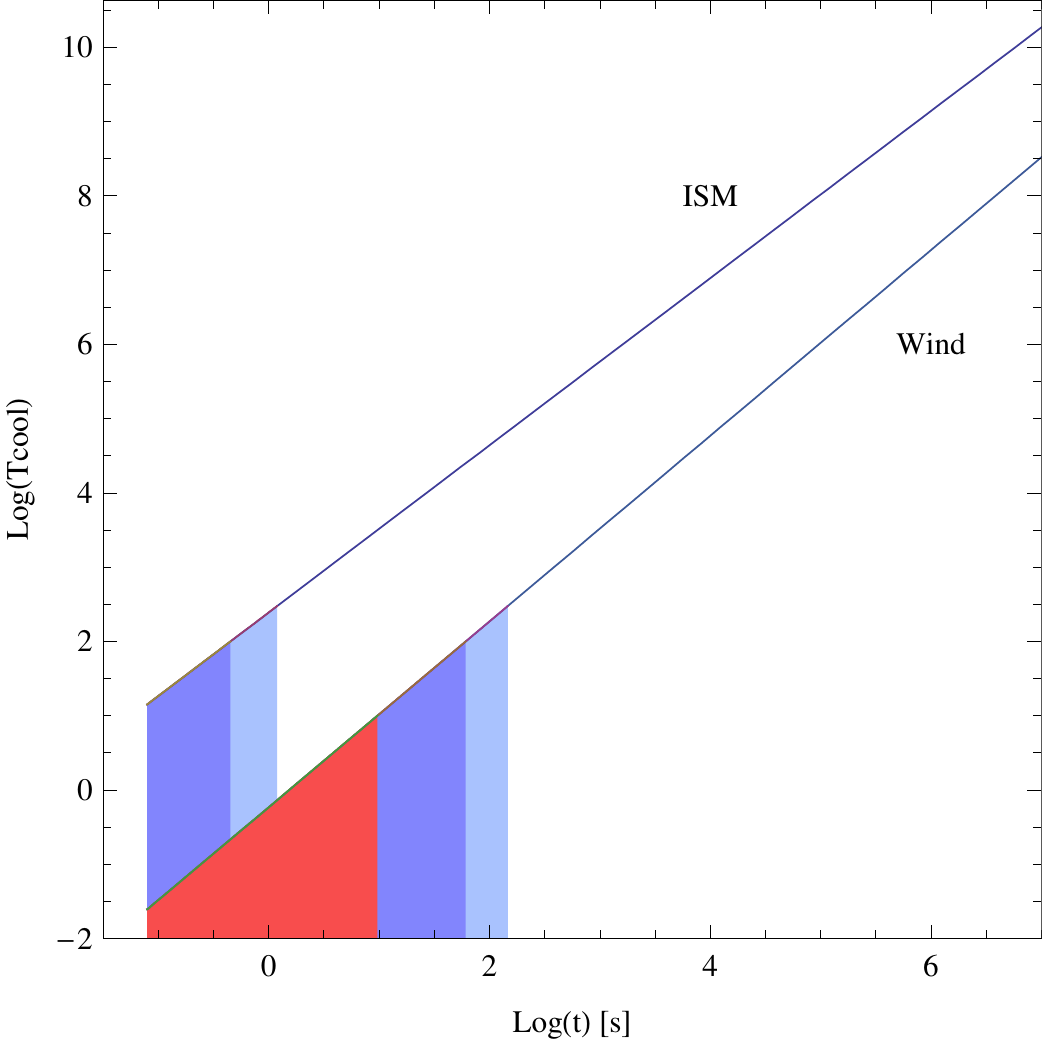}
\caption{Cooling time in afterglows vs. time after the burst, for the ISM
and Wind models of the external density. In the left panel, we use the 
``typical'' parameters $E=10^{53}$~erg, $A_*=1$ and 
$n_{\rm ISM}=1~{\rm cm}^{-3}$,
whereas in the right panel, we use a rather extreme set of parameters, 
$E=10^{54}$~erg, $A_*=10$ and $n_{\rm ISM}=100~{\rm cm}^{-3}$. In both cases, 
we put a GRB at a typical $z=2$; time in the plots ranges 
from 0.1~s to 100~days. Color coding is the same as in figure \ref{f1}.
\label{f4}}
\end{figure}

The most remarkable results are obtained for internal shocks in the
baryon-dominated scenario. The ``diagrams of state'' -- the contours of 
constant $T_{\rm cool}$ in the planes $R$-$\Gamma$ and $R$-$\Gamma_i$
are shown in figures \ref{f1}, \ref{f2}. One can see that the regime of 
{\it strong cooling} is limited to low bulk Lorentz factors, $\Gamma\la60$ 
and small radii $R\la10^{10}-10^{11}$~cm. Such shocks are well 
below the photosphere, located at $R_{\rm ph}\sim10^{14}-10^{15}$~cm
for such low $\Gamma$'s, which makes them hardly observable. 
However, {\it radiative shocks} with $T_{\rm cool}\sim100-300$ can form
above the photosphere for the conventionally assumed GRB parameters,
$\Gamma\sim200-300$, $L\sim10^{52}$~erg/s, $\Gamma_i\ga2.5$, as
is seen from figure \ref{f2}. In this figure, we considered shocks
forming at the photospheric radius, i.e., when 
$R_i=R_0(2\Gamma^2)\simeq R_{\rm ph}$. One should keep in mind that 
internal shocks can occur at radii smaller than $R_i$ if the outflow
modulation occurs due to instabilities in the outflow/jet itself,
at radii $R\gg R_0$ where the flow is already ultra-relativistic, 
rather than at the base of the jet, at $R\sim R_0$ \citep{RM05}.
This substantially relaxes the conditions on the optical depth, 
especially at high $\Gamma$. Note that depending on the efficiency of 
conversion of the kinetic energy into radiation, the radiation from
internal shocks can produce the second, $e^\pm$-pair photosphere.
Should this happen, the optical depth is estimated to be from few to ten.
One can expect, therefore, the appearance of a thermal component in the
spectrum. Such optical depths are, nevertheless, not enough to 
completely smear out a non-thermal component; $\tau_\pm$ 
greater than a hundred are likely needed for this \citep{PW04}.  
The pair opacity is less of a problem for low-energy GRBs: 
for $L\sim10^{48}-10^{49}$~erg/s, $\Gamma\sim200$ 
and the radiation efficiency of about 10\%,
the radiative shocks can occur at $\tau_\pm\sim1$
and well above the baryonic photosphere.   

The existence of the regime of GRBs with strong radiative cooling has 
many interesting implications. 

The first of them concerns  radiation 
emitted at such shocks. The magnetic fields at the shock front are 
highly inhomogeneous and anisotropic on very small scales, much 
smaller than the typical electron gyro-radius.
The fields predominantly have the filamentary structure, 
reminiscent of the filaments of the
Weibel instability undergone subsequent numerous mergers
in the foreshock region (see, e.g., \citealp{Spit08,CSA08}).
It has been predicted that radiation produced in such small-scale fields
(called the ``jitter radiation'') should be spectrally different 
from the conventional synchrotron radiation \citep{M00}. The steeper
than synchrotron low-energy spectral slope, $F_\nu\propto\nu^1$,
was an attractive solution to the ``synchrotron line of death'' problem 
in prompt GRBs \citep{Preece+98,M00}. Because of the anisotropic, filamentary 
structure of magnetic fields at the shock front, jitter radiation 
is predicted to be anisotropic too, with it's spectral shape varying
with angle between the shock normal and the line of sight \citep{M06}. 
Combined with relativistic kinematics, the model reproduces certain 
spectral correlations observed in the time-resolved analysis of GRB data
\citep{Kaneko+06}, as well as the ubiquity of flat, $F_\nu\propto\nu^0$, 
spectra in the sample \citep{Poth+07}. 
Afterglow lightcurves in the jitter regime have also been recently 
calculated \citep{M+07,Workman+07}. 
 
The second one deals with the fact that realistic 
GRB shocks in a certain parameter regime are now accessible to simulations. 
Present 2D PIC simulations
already resolve the shock formation and evolution, generation of
magnetic fields in the foreshock region and their amplification toward 
the shock front, the electron acceleration to near equipartition
with the protons and, tentatively, Fermi-type acceleration.
The shocks in the radiative regime are remarkable in that the energetic 
electrons are able to radiate a substantial fraction of the shock 
energy, thus making the shocks radiatively efficient and, hence, observable
as prompt GRBs. This opens the possibility to ultimately 
understand the properties of collisionless relativistic shocks.
Moreover, by adding radiative cooling in the PIC simulations,
one can study to what extent rapid cooling of electrons affects the shock
structure on the microscopic scale. At last, one 
can obtain the radiation spectra directly from PIC simulations
(this possibility has already been demonstrated by \citealp{Hededal05}),
compare them with observational data and, after all,
confirm or falsify the jitter model of prompt GRBs.
All this does not seem feasible at present, however, for the shocks in 
the weakly radiative regime, in which the structure and dynamics 
of the magnetic field far behind the shock is crucial, but which is 
still difficult to trace with available computational resources.

Third, in the radiatively efficient regime, an electron loses its energy every time it is crossing the shock, so Fermi acceleration of the electrons will be quite inefficient. One can expect that a power-law electron distribution will not form in this case and, as a consequence, the observed radiation spectrum may not have an extended non-thermal tail. Moreover, the peak of the emitted radiation is predicted for such a regime to be in the multi-MeV range, where the comptonized component is conventionally expected, but no strong second peak is expected in the keV-MeV range. This prediction can soon be tested with {\it Fermi} (former {\it GLAST}) observatory. 

Finally, the Weibel instability and radiation production in conditions 
very close to those in GRBs can be studied in existing laboratory experiments. 
The generation of filamentary magnetic fields,
indicative of the Weibel instability, has been demonstrated
in numerous laser-produced plasma experiments (see, e.g., \citealp{Tatar+03}).
In these experiments, a Petawatt laser beam with the intensity
$I\sim10^{20}-10^{22}~{\rm W~cm}^{-2}$ produces a relativistic plasma
of density $n\sim10^{19}-10^{21}~{\rm cm}^{-3}$ and relativistic electrons 
with the Lorentz factor of one to few hundred. Because the
the density in the experiments is so close to the density in internal shocks, 
$\sim10^{16}~{\rm cm}^{-3}$, the plasma skin depths differ by about
two orders of magnitude, cf., $c/\omega_{pp}\sim10~\mu$m in the lab and
$\sim{\rm few~mm}$ in prompt GRBs. For the typical transverse size
of a laser beam of $\sim100~\mu$m, a large number of the Weibel filaments 
are produced in the target, which makes the further study of the 
nonlinear dynamics of filaments feasible.
By launching a probe relativistic electron beam through the target and
performing the X-ray and beam diagnostics, as is proposed in the 
{\it Hercules} 
experiment\footnote{http://www.eecs.umich.edu/CUOS/research/index.html}, 
one can directly and simultaneously probe the magnetic field structure 
and the spectral properties of the emitted radiation.
These experiments can be done in a number of existing laser
facilities, such as {\em Vulcan, Hercules, Omega, NIF} and a few other. 
Apart from being a very exciting possibility for the
{\it Laboratory Plasma Astrophysics}, such experimental studies
are crucial for the verification of numerical results.   

In this paper, we evaluated the cooling time in GRBs with reasonable accuracy.
One should understand, however, that there are intrinsic uncertainties
in the values of certain parameters, which are difficult to
quantify. In our calculation of the average residence time of an 
electron in a region of high magnetic field at the shock front,  
we did not take into account the effective filling factor 
of the field, nor did we consider the increase of the residence
time due to the electron trapping. Such effects strongly depend
on individual particle trajectories and are, therefore, sensitive to
both the field structure and the particle distribution function.
We, however, evaluated and included the effect of the clumpiness 
of the magnetic field on the effective (emission-weighted) 
$\hat{\epsilon}_B$, as it is relatively straightforward. 
Here we also used the results of 2D simulations, which may  
differ from a realistic 3D case.
More importantly, in the strong cooling regime, the electron
cooling occurs on the time-scale of the shock formation.
This effect can, in principle, change the entire shock structure.
Since the effect of radiative cooling has never been studied before,
the results of this paper are suggestive.
How does strong electron cooling change the collisionless shock 
formation, its structure and dynamics? The answer to this question 
can be obtained from future numerical PIC simulations which include
radiative effects. However, we can speculate that shocks with strong electron cooling will likely have longer transition regions than the non-radiative shocks. This is because electron heating is an integral part of the electron-ion shocks \citep{Spit08}, as it enables electrons to escape the ion filaments. This  reduces filament shielding, and facilitates ion filament merging, which leads to shock formation. If electrons are kept cold in the upstream, the shock may thus take longer to form \citep{LyubEich07, Ged08}. 
Finally, in our study we completely ignored the magnetic fields present in the upstream region. These fields are relatively week, compared to the fields at the shock itself, but they can make a non-negligible contribution to the electron cooling if they occupy a large volume. Simulations of electron-positron shocks indicate that even in the longest simulations the upstream magnetic field has not reached a steady state, but continues to grow with the simulation time \citep{KKSW08}. In our analysis we also ignored the presence of the fields in the downstream, which have also been demonstrated to be affected by the upstream fields. Thus, our analysis is quite conservative since both the upstream and downstream fields can only decrease the electron cooling time and/or increase the radiative efficiency of the shock.

\ack
This work has been supported by NSF grants AST-0708213, AST-0807381, NASA ATFP grant NNX-08AL39G, Swift Guest Investigator grant NNX-07AJ50G and DOE grant  DE-FG02-07ER54940. A.S. acknowledges the support from Alfred P. Sloan Foundation
fellowship.

\end{document}